\documentstyle[emulateapj,psfig,apjfonts]{article}
\def\g328{G328.4+0.2}
\def\etal{{\rm et~al.\ }}
\def\HI{H\,{\sc i}}
\def\HII{H\,{\sc ii}}
\def\kms{km~s$^{-1}$}

\lefthead{GAENSLER, DICKEL AND GREEN}
\righthead{\g328: A LARGE AND LUMINOUS CRAB-LIKE SNR}

\begin{document}
\title{\g328: A large and luminous Crab-like supernova remnant}
\submitted{(To appear in {\em The Astrophysical Journal})}
\author{B. M. Gaensler\altaffilmark{1,5}, J. R. Dickel\altaffilmark{2,3}
and A. J. Green\altaffilmark{4}}

\altaffiltext{1}{Center for Space Research, Massachusetts Institute of
Technology, 70 Vassar Street, Cambridge, MA 02139; bmg@space.mit.edu}
\altaffiltext{2}{Astronomy Department, University of Illinois, 1002 West
Green Street, Urbana, IL 61801; johnd@astro.uiuc.edu}
\altaffiltext{3}{Netherlands Foundation for Research in
Astronomy, PO Box 2, 7990 AA Dwingeloo, The Netherlands}
\altaffiltext{4}{Astrophysics Department, University of Sydney,
NSW 2006, Australia; a.green@physics.usyd.edu.au}
\altaffiltext{5}{Hubble Fellow}

\begin{abstract}

We report on radio continuum and \HI\ observations of the radio source
\g328\ using the Australia Telescope Compact Array. Our results confirm
\g328\ to be a filled-center nebula with no surrounding shell, showing
significant linear polarization and an almost flat spectral index.
These results lead us to conclude that \g328\ is a Crab-like, or
``plerionic'', supernova remnant (SNR), presumably powered by an unseen
central pulsar.  \HI\ absorption towards \g328\ puts a lower limit on
its distance of $17.4\pm0.9$~kpc, making it the largest ($D=25$~pc) and
most luminous ($L_R = 3\times10^{35}$~erg~s$^{-1}$) Crab-like SNR
in the Galaxy. We infer \g328\ to be significantly older than the
Crab Nebula, but powered by a pulsar which is fast spinning
($P<20$~ms) and which has a comparatively low magnetic field ($B<10^{12}$~G).
We propose \g328, G74.9+1.2 and N157B as a distinct group of
large-diameter, high-luminosity Crab-like SNRs, all powered by
fast-spinning low-field pulsars.

\end{abstract}


\section{Introduction}
\label{sec_intro}

When a massive star ends its life in a supernova explosion, the outer
layers of the star are expelled at high velocity, and interact with the
ambient interstellar medium (ISM) to produce a supernova remnant (SNR).
``Shell'' SNRs are usually characterized by radio synchrotron emission
with a limb-brightened morphology, roughly centered on the site of
the supernova explosion.  Some SNRs are classified as ``composite'',
indicating that as well as a shell they have an additional central component,
characterized by a filled-center morphology, a flat spectral index
($-0.3 \la \alpha \la 0$; $S_\nu \propto \nu^\alpha$) and
significant linear polarization. This extra component is usually
interpreted as a synchrotron nebula, 
powered by a pulsar also formed in the supernova
explosion (\cite{mgh+79}; \cite{wp80}; \cite{rc84}).  Thus even when a
pulsar itself is not detected (as is the case much more often than not),
the mere existence of one of these synchrotron nebulae tells us that an
energetic pulsar must be located within.

The best known example of a pulsar-powered nebula is the Crab Nebula.
However, it has long been apparent that the Crab is very different
from most other SNRs, in that it consists of a synchrotron nebula and
associated pulsar, but has {\em no}\ surrounding shell corresponding
to the supernova blast wave.  It is now recognized that about 5\% of
all SNRs similarly lack shells; these sources are generally referred
to as ``Crab-like'' SNRs, or ``plerions'' (\cite{wp78}). It is still
an open question as to whether Crab-like SNRs have surrounding shells
which are simply not detectable (\cite{che77}; \cite{sh97}), or whether
they are fundamentally different from other SNRs in that they
have no shell, invisible or otherwise (\cite{nom87}; \cite{wlt97}).  

Crab-like SNRs are thus important to identify and study, both for
understanding the nature and evolution of SNRs,
and because they are an unambiguous indication of the presence of a
young pulsar.  The source \g328\ (MSH~15--5{\em 7}; \cite{msh61})
has been been the subject of several radio studies (\cite{sg70c};
\cite{chmw80}; \cite{wg96}), which have indicated a
featureless filled-center appearance and a comparatively flat spectral
index. These properties have resulted
in \g328\ being proposed as a member of the Crab-like
class of SNRs, and indeed is
the only Galactic Crab-like SNR easily accessible to
Southern hemisphere observers.  \HI\ measurements with a two-element
interferometer have shown \HI\ absorption in this direction to a distance
in excess of 20~kpc (\cite{cmr+75}). This large distance would make the
SNR extraordinary luminous, even brighter than the Crab Nebula itself.

However, previous images of this source have been of only intermediate
resolution (the best observations to date had only six beams across the
SNR), while the lack of imaging capability in the \HI\ measurements of
Caswell \etal\ (1975\nocite{cmr+75}) mean that the absorption
they saw could have been coming from another bright source in
the vicinity, rather than from the SNR itself.
We therefore present new radio observations of \g328\ both in polarimetric
continuum and in the \HI\ line, aimed at confirming the Crab-like
nature of this source, verifying the large distance claimed for it, and
studying it at much higher resolution and sensitivity than in previous
work. Our observations and analysis are described in \S\ref{sec_obs},
and our results are presented in \S\ref{sec_results}. In
\S\ref{sec_discuss} we confirm \g328\ as a Crab-like SNR, and infer the
properties of the central pulsar presumed to be powering it.
A complementary study of this source in X-rays is
reported in a companion paper by Hughes, Slane \&
Plucinsky (2000\nocite{hsp00}).

\section{Observations and Data Reduction}
\label{sec_obs}

Our observations of \g328\ were made with the Australia Telescope Compact
Array (ATCA; \cite{fbw92}), 
a 6~km east-west synthesis array located near Narrabri,
NSW, Australia, on dates as listed in Table~\ref{tab_obs}. Continuum
observations were made at 1.4 and 4.5~GHz; each band had a width of
128~MHz, divided into 32 channels. Simultaneous with the 1.4~GHz data,
observations were made in the \HI\ line, centered on 1421~MHz and
using 512 channels across a 4~MHz bandwidth. All four Stokes parameters
({\em XX}, {\em YY}, {\em XY} and {\em YX}) were recorded in continuum
observations; only {\em XX} and {\em YY} were obtained in the \HI\ line.
All observations consisted of a single pointing centered on \g328. Flux
density calibration was carried out using observations of PKS~B1934--638
(with assumed flux densities of 14.9 and 6.3~Jy at 1.4 and 4.5~GHz
respectively), while antenna gains and polarization were calibrated using
regular observations of MRC~B1456--367 at 1.4~GHz and PMN~J1603--4904
at 4.5~GHz.

Data were reduced in the {\tt MIRIAD}\ package. After flagging
and calibration were carried out, total intensity images
were formed at each of 1.4 and 4.5~GHz using uniform weighting,
multi-frequency synthesis and maximum entropy deconvolution.
The resulting images were then corrected for the mean primary
beam response, with resolutions and sensitivities given
in Table~\ref{tab_obs}.

Images in linear polarization were formed by creating separate
images in Stokes $Q$ and $U$, deconvolving each of these
separately using the {\tt CLEAN}\ algorithm, and then combining
with appropriate debiasing to form $L = (Q^2 + U^2)^{1/2}$.
At 1.4~GHz, significant Faraday rotation across the band can
depolarize the emission, and images of linear polarization were
produced on a channel by channel basis and then averaged
to form a single map. At 4.5~GHz, differential Faraday effects are
negligible and the entire observing bandwidth was used to form
a single pair of Stokes $Q$ and $U$ images. The resulting
maps of linear polarization were clipped wherever
Stokes $I$, $Q$ or $U$ fell below 3$\sigma$.

For the \HI\ data, the continuum contribution was subtracted
in the {\em u-v} plane (\cite{vc90}),
and a cube then formed using
uniform weighting and discarding all baselines longer than
7~k$\lambda$. The cube
contained planes between --200~\kms\ and +200~\kms\ (LSR)
at intervals of 4~\kms. The peak flux density in the image was sufficiently
low that no sidelobes were apparent, and no attempt was made to
deconvolve. The cube was then weighted by the corresponding 1.4~GHz
continuum image. Absorption spectra against SNR~\g328\ and
the nearby \HII\ region G328.30+0.43 were then
generated by integrating over an appropriate spatial region and
renormalizing appropriately to give units of fractional absorption.

The rms noise, $\sigma$, in each absorption spectrum was estimated
from the fluctuations in line-free channels. As in previous papers
(e.g. \cite{gmg98}),
we adopt $6\sigma$ as a threshold for significance,
to take into account the increase in system temperature of the receivers
due to the brightness temperature of \HI\ in the Galactic Plane.

\section{Results}
\label{sec_results}

\subsection{Continuum}
\label{sec_results_cont}

\begin{figure*}[htb]
\centerline{\psfig{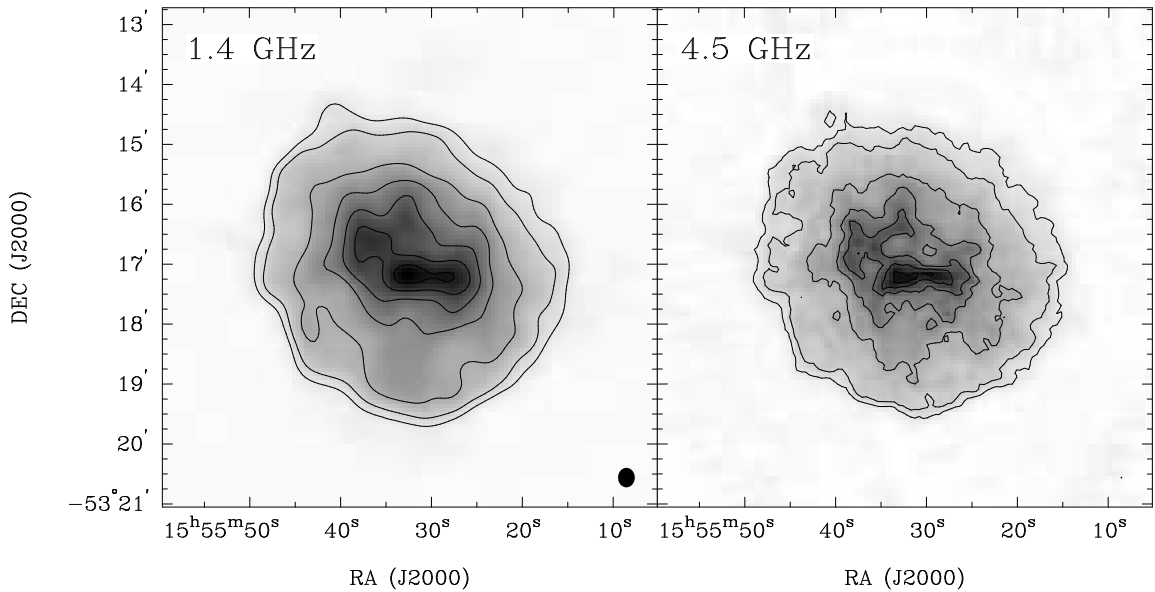}}
\caption{1.4 and 4.5~GHz ATCA images of SNR~\g328, corrected
for primary beam attenuation. The 1.4~GHz
image has a greyscale which ranges
from --5 to 200~mJy~beam$^{-1}$, with contour levels
at 15, 30, 60, $\ldots$, 150, 180~mJy~beam$^{-1}$.
The 4.5~GHz image ranges from --0.05 to
2.0~mJy~beam$^{-1}$, with contours 
at levels of 0.15, 0.30, 0.60, $\ldots$, 1.50, 
1.80~mJy~beam$^{-1}$. The synthesized beam is shown at the
lower right of each panel.}
\label{fig_snr}
\end{figure*}

Images of \g328\ at 1.4 and 4.5~GHz are shown in Fig~\ref{fig_snr},
and properties of the source at each frequency are given in
Table~\ref{tab_obs}.  The source is approximately circular, with a bar-like
feature running east-west at the peak of the emission. A plateau of
fainter emission surrounds this bar; this plateau is significantly more
pronounced to the east and to the south.

\g328\ is of sufficiently small diameter
that its entire flux density is recovered by the telescope
at both observing frequencies, and the corresponding flux
densities (after a background correction has been applied)
are given in Table~\ref{tab_obs}.
These flux densities, together with the 0.8~GHz measurement
of Whiteoak \& Green (1996\nocite{wg96}), imply a spectral
index for the source $\alpha = -0.12\pm0.03$.

Examination of the whole 1.4~GHz field (not shown
here) shows there to be no outer shell of emission surrounding \g328,
out to a radius of 15~arcmin and down to a 1$\sigma$ sensitivity of
0.8~mJy~beam$^{-1}$, corresponding to a surface brightness limit
at 1~GHz (assuming a typical shell spectral index $\alpha = -0.5$)
of $\Sigma= 1.6 \times 10^{-21}$~W~m$^{-2}$~Hz$^{-1}$~sr$^{-1}$.

Discarding the shortest baselines to filter out
extended emission, we can put $5\sigma$ limits at 1.4~GHz and 4.5~GHz
of 7 and 0.3~mJy respectively on the flux density of any point source
within \g328.

\subsection{Polarization}

At 1.4~GHz, the only linearly polarized emission seen from \g328\ is in
a couple of clumps seen at the position of the peak in total intensity.
These clumps have a fractional polarized intensity of $\sim$1\%.  When
imaged on a channel by channel basis, the variation of the position angle
of this polarization, $\phi$, shows a $\phi \propto \lambda^2$ dependence
across the 1.4~GHz band.
This allows us to determine a rotation measure (RM)
for this emission (cf. \cite{gmg98}) of $-900\pm100$~rad~m$^{-2}$, where
the uncertainty is dominated by spatial variations in RM rather than by
uncertainty in the measurement process (for which typical uncertainties
are $\pm30$~rad~m$^{-2}$).

\begin{figure*}[htb]
\centerline{\psfig{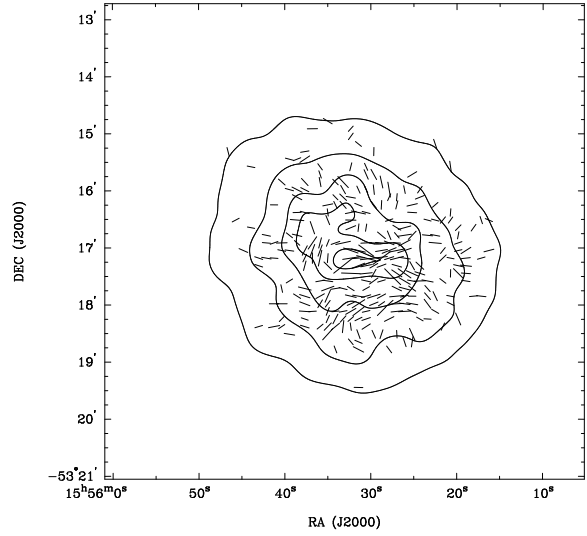}}
\caption{Linearly polarized emission from \g328\ at 4.5~GHz.
The length of each vector represents the polarized intensity
at that position, while the position angle corresponds to the
orientation of the electric field vector when averaged across
the entire 4.5~GHz band. The longest vector in the image
is for a polarized intensity of 0.6~mJy~beam$^{-1}$.
The contours represent total intensity
at 4.5~GHz, smoothed to a resolution of $20''$.}
\label{fig_pol}
\end{figure*}

At 4.5~GHz, significantly more linear polarization is seen from
\g328, with a spatial distribution shown in Fig~\ref{fig_pol}. Although
the polarization appears to be reduced around the edges of the source,
this is most likely an artifact of the clipping of polarized emission
below $3\sigma$ in Stokes~$Q$ and $U$; the {\em fractional} polarized
intensity shows no decrease towards the edges. The level of
fractional polarization ranges between 5\% and 50\% with a mean of 20\%.

An RM of 1000~rad~m$^{-2}$ corresponds to $\Delta \phi \approx 10^\circ$
across the entire 4.5~GHz observing bandwidth, which is smaller than
the uncertainties in position angles of individual frequency channels.
Thus no rotation measures could be extracted from the 4.5~GHz data.

\subsection{\HI\ absorption}

\begin{figure*}[htb]
\centerline{\psfig{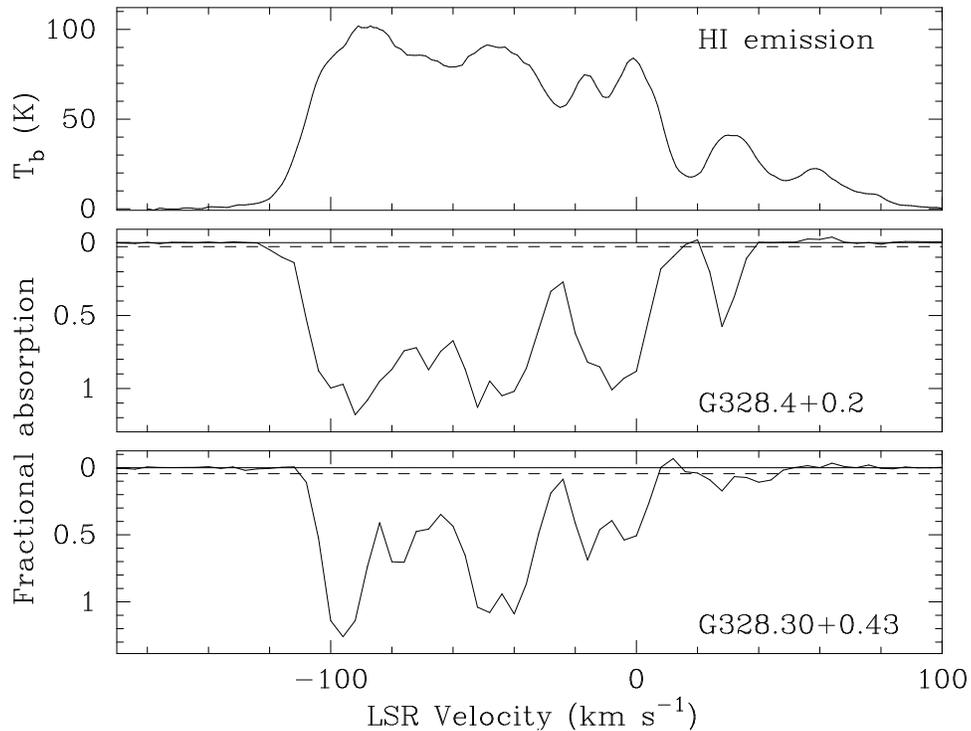}}
\caption{\HI\ emission and absorption in the region. The upper panel
shows the \HI\ emission spectrum measured by 
Kerr \etal\ (1986)\protect\nocite{kbjk86} in the direction $l = 328.5$,
$b = 0.25$. The central panel shows \HI\ absorption 
towards SNR~\g328, while the lower panel shows absorption
towards the nearby \HII\ region G328.30+0.43. The dashed lines
in the absorption profiles
correspond to absorption at levels of 6$\sigma$, where
the rms has been computed from line-free channels.}
\label{fig_hi}
\end{figure*}

\HI\ absorption towards \g328\ needs to be compared to an
\HI\ emission spectrum in a similar part of the sky. But
because most of the power in \HI\ emission is on scales larger
than were sampled by our observations, such a
spectrum is difficult to extract from our ATCA data.
Instead, we use an \HI\ emission spectrum taken from
the single-dish survey of Kerr \etal\ (1986)\nocite{kbjk86},
in the direction $l = 328\fdg5$, $b=0\fdg25$. This emission profile
is shown in the upper panel of Fig~\ref{fig_hi}.
Two ATCA absorption profiles are also shown in Fig~\ref{fig_hi}: towards
\g328\ and towards the \HII\ region G328.30+0.43, 15~arcmin to the
north-west. The latter emits in various maser lines, and is known
to have a systemic velocity of approximately --95~\kms\ 
(\cite{be83}; \cite{cmc95}).

At negative velocities, the two spectra are very similar, and match
well the \HI\ emission seen along similar lines-of-sight.  However, a
significant difference is seen at positive velocities, where \HI\ at a
velocity of +28~\kms\ is clearly producing significant absorption against
\g328, but only a weak feature is seen for G328.30+0.43.  Examination of
the image plane corresponding to this velocity clearly shows the outline
of all of \g328\ in absorption, leaving no doubt that this absorption
feature is genuine. However, for the \HII\ region G328.30+0.43 we see
no real match in absorption to the source's morphology at this velocity,
as expected given that it is known to be at the tangent point.
This weak absorption probably corresponds to fluctuations in the \HI\
emission at this velocity, appearing negative because of spatial filtering
by the interferometer.

\section{Discussion}
\label{sec_discuss}

\subsection{Distance}
\label{sec_dist}

The \HI\ spectrum shown in Fig~\ref{fig_hi} demonstrates clearly that
the systemic velocity of \g328\ is at least +28~\kms.  \HI\ emission is
seen at +28~\kms\ not just in the direction of \g328, but towards
several other sources in the vicinity (\cite{cmr+75}). Thus it is
unlikely that the emission and absorption seen at this velocity
correspond to some much closer cloud of gas which deviates dramatically
from the Galactic rotation curve.  While an \HI\ emission feature at
+58~\kms\ shows no corresponding absorption, its low brightness
temperature means that we can probably not use this result to derive an
upper limit on the source's radial velocity.

To calculate a lower limit on the distance to \g328, we adopt standard
IAU parameters for the Sun's orbital velocity ($\Theta_0 = 220$~\kms)
and distance from the Galactic Centre ($R_0 = 8.5$~kpc) (\cite{klb86}),
use the best fitting model for Galactic rotation of Fich, Blitz \&
Stark (1989\nocite{fbs89}), and assume uncertainties in systemic
velocities of $\pm$7~\kms. For a systemic velocity of +28~\kms, it then
follows that \g328\ must be at a distance of at least $17.4\pm0.9$~kpc.
In further discussion, we assume a distance $d = 17d_{17}$~kpc; the
diameter of \g328\ is then $25d_{17}$~pc.

\subsection{Morphology}

\g328\ is an extended Galactic source with a filled-center morphology,
significant linear polarization across most of its extent, and a spectral
index much flatter than for a typical shell-type supernova remnant. Down
to our surface brightness limit,  there is no evidence that \g328\ is
surrounded by a larger, limb-brightened shell. This limit is sufficient
to detect the shell of a typical young SNR (see \cite{fkcg95}),
although not sensitive enough to detect the shell component of the
composite SNR~G322.5--0.1 (\cite{whi92}). Thus while the limits on a
shell around \g328\ are not as stringent as around some other such sources,
from the available data we class it as a Crab-like SNR.

The mere presence of a Crab-like or composite
SNR is usually taken to indicate that there
is an energetic pulsar within, even though in most such sources no
pulsar has yet been detected. While indeed no pulsar has been seen
within \g328\ in either radio waves (\cite{mdt85}; \cite{kmj+96}) or in
X-rays (\cite{wil86}; \cite{hsp00}),
this can easily be accounted for by unfavorable
beaming, insufficient sensitivity at the SNR's large distance, and significant
scattering, dispersion (at radio wavelengths) and absorption (in X-rays)
along the long line of sight. We thus assume in further
discussion that there is an unseen rotation-powered pulsar within \g328,
whose relativistic wind powers the radio emission from the remnant.

The bar running through the center of \g328\ is a feature similar to that
seen in many other Crab-like SNRs. There are two interpretations of such
a feature: one is that it corresponds to a trail of emitting particles
left behind by a pulsar with a significant space velocity (e.g.\ 
\cite{fggd96}; \cite{swc99}),
the other is that the bar is actually comprised of
two opposed jet-like features, produced by a bipolar outflow from the
pulsar (e.g.\ F\"{u}rst \etal\ 1988, 1989\nocite{fhm+88,fhr+89}). 
High-resolution X-ray imaging is
required to distinguish between these possibilities, since the much shorter
synchrotron lifetime of X-ray emitting particles will result in X-ray
emission being concentrated around the current point of injection into the
SNR. So if the bar is a high velocity pulsar leaving a trail, we expect
to see X-ray emission concentrated at one end of the bar, while if it is
a pair of outflows, X-ray emission should peak near the center of the bar.

\subsection{Physical Properties}

We first determine the radio luminosity, $L_R$, of \g328.
Hughes \etal\ (2000\nocite{hsp00}) demonstrate that
any break in the continuum spectrum of this source
is at a frequency higher than 100~GHz. We thus
assume that the radio spectrum of \g328\ is a
single power-law of spectral index $\alpha = -0.12$.
Integrating between 100~MHz and 100~GHz, we then
find a broad-band radio
luminosity for \g328\ of
$L_R = 3.3d_{17}^2\times10^{35}$~erg~s$^{-1}$.  

In Table~\ref{tab_compare}, we compare the size and radio luminosity of \g328\
to those of other Crab-like SNRs (as well as Crab-like cores of composite
SNRs) with similar properties; it can be seen that
\g328 is one of the most radio-luminous and largest Crab-like
SNRs yet discovered. Two sources closely resemble \g328:
N157B (\cite{wg98b};
\cite{ldh+00}), a Crab-like SNR in the Large Magellanic Cloud (LMC) in which
a 16-ms X-ray pulsar has recently been discovered\footnote{The pulsar has
so far remained undetected in radio waves, but otherwise
has properties typical of a radio pulsar (\protect\cite{ckm+98}).},
and G74.9+1.2 (\cite{ws78}; \cite{wltp97}), a 
Galactic SNR which as yet has had no pulsar detected in it. Below we will
argue that these three SNRs constitute a particular
sub-class of Crab-like SNR.

A pulsar's spin-down luminosity is given by 
\begin{equation}
\label{eqn_edot}
\dot{E} \equiv 4I\pi^2 \dot{P}/P^3,
\end{equation}
where $P$ is the period of the pulsar and  $I = 10^{45}$~g~cm$^{-2}$
is its moment of inertia.
Defining $\epsilon = L_R/\dot{E}$, we find that
in the few cases where a pulsar has been detected within a Crab-like SNR,
the radio luminosity of the SNR normally falls in the narrow range
$\epsilon = (1-5)\times 10^{-4}$ (\cite{fs97}; \cite{gsf+00}).
In particular, the Crab Nebula has $\epsilon = 4\times10^{-4}$, and
we assume a similar value for \g328. One then finds that its
central pulsar has $\dot{E}_{38}=8.3$
(where $\dot{E} = 10^{38}\dot{E}_{38}$~erg~s$^{-1}$), a value
consistent with that derived by Hughes \etal\ (2000\nocite{hsp00})
from the X-ray properties of this source.
This value of $\dot{E}$ ranks
amongst the highest values seen for the radio pulsar population,
and makes this source a good candidate for pulsed emission searches
by future X-ray and $\gamma$-ray missions.

A pulsar drives a 
supersonic bubble into its surroundings, whose radius,
$R$, is given by (\cite{wmc+77}; \cite{aro83}): 
\begin{equation} 
R = 0.82 \dot{E}_{38}^{1/5} t_3^{3/5} n_0^{-1/5}\, {\rm pc}, 
\label{eqn_arons} 
\end{equation} 
where $t_3$~kyr is the age of the pulsar and $n_0$~cm$^{-3}$ is the
density of the ambient medium. It is usually assumed
that Crab-like SNRs are propagating into a low density component of
the ISM (\cite{che77}), for which we adopt a typical density 
$n_0 = 0.003$~cm$^{-3}$.
We can then determine an age
for \g328\ of $t_3 \approx 7d_{17}^{5/3}$~kyr.

For a idealized braking index $n=3$ and an initial period $P_i \ll P$,
the age of the SNR is equal to its pulsar's
characteristic age, $\tau_c \equiv P/2\dot{P}$. We can then combine our
inferred age with Equation~(\ref{eqn_edot}) to determine parameters
for the pulsar of $P=11$~ms
and $B \equiv 3.2\times10^{19} (P\dot{P})^{1/2}$~G~$= 6 \times 10^{11}$~G, where
$B$ is the inferred surface magnetic field of the neutron star,
assuming a dipole geometry. 
These inferred parameters, along with the results of similar calculations
for G74.9+1.2 and N157B, are given in Table~\ref{tab_compare} -- we
find that all three pulsars are of much smaller period,
lower magnetic field and larger age than the Crab Pulsar.

Various assumptions have gone into the calculations
behind this last statement, and we now
consider what effect each one has on the results.  The assumption least
likely to be true is that $t=\tau_c$; if we relax this requirement to
include braking indices $n \neq 3$ and initial periods $P_i \approx P$,
we still find, for a wide range of parameters, that $\tau_c \ga 0.7 t$
(see Fig~3 of \cite{mgz+98}). Since $P \propto \tau_c^{-1/2}$ and $B
\propto \tau_c^{-1}$, the consequent inferred spin periods and magnetic
fields are still significantly smaller than for the Crab Pulsar.  A
second assumption is that $\epsilon = \epsilon_{\rm Crab}$; however the
dependence on $\epsilon$ is even weaker than on $\tau_c$ ($P \propto
\epsilon^{1/3}$ and $B \propto \epsilon^{1/6}$),  so that even invoking
a value of $\epsilon$ ten times larger than that for the Crab (which
would be higher than for any pulsar observed) has little effect on the
results.  We have also assumed a value $n=0.003$~cm$^{-3}$; a significantly
lower density than this can indeed produce a more typical magnetic
field ($B \propto n^{-2/3}$), but still requires a fast period ($P
\propto n^{-1/6}$).  Meanwhile,
Equation~(\ref{eqn_arons}) is only
strictly true for constant $\dot{E}$. In reality, $\dot{E}$ will
decrease as a function of time, so that the radius of the pulsar wind
bubble will be larger than given by Equation~(\ref{eqn_arons}) and will
hence cause us to over-estimate the age of the system. While this
effect is difficult to quantify, the ages given in Table~\ref{tab_compare}
would have to be too large
by a factor of 5--10 to produce periods as slow as that of
the Crab Pulsar. Finally, we note that the calculations we have made 
for \g328\ all use the lower limit on the distance 
derived in Section~\ref{sec_dist}; at greater
distances, the inferred period and magnetic field are both even lower
than those given in Table~\ref{tab_compare}.

Thus even if not all of the assumptions which we have made are valid,
one is still forced to conclude that the pulsars powering these SNRs
are likely to be spinning rapidly and have low magnetic fields.
In particular, the pulsar actually observed within N157B indeed fulfils
this prediction, having observed properties quite similar to those we
have inferred for it (see Table~\ref{tab_compare}).

\section{Conclusions}

Our radio observations of \g328\ confirm it to be a Crab-like SNR at a
distance of $>$17~kpc, making it the largest and most radio-luminous such
object in our Galaxy. \g328, together with the Galactic SNR G74.9+1.2 and
N157B in the LMC, appear to form a small subset of Crab-like SNRs with
both high radio luminosities and large diameters.  The high luminosities
of these remnants demand a high value of $\dot{E}$ for their central
pulsars, while these remnants' large extent, even assuming an energetic
pulsar and a low ambient density, requires them to be significantly
older than the Crab Nebula.  Since $\dot{E}\tau_c \propto P^{-2}$ and
$\dot{E}\tau_c^2 \propto B^{-2}$, this combination of a high $\dot{E}$
and large age can only be produced by a pulsar which is spinning
at least twice as fast as the Crab Pulsar, 
but which has a magnetic field $\sim$5
times weaker.  
Specifically, we infer \g328\ to be
$\sim$7000~yrs old and powered by a pulsar with period $P\approx11$~ms
and dipole magnetic field $B=6\times10^{11}$~G. 
Similar properties are
predicted for G74.9+1.2 and for N157B, which in the latter case agree
with those of the pulsar recently detected within this SNR.

Using the distribution of pulsar initial magnetic fields proposed by
Stollman (1987\nocite{sto87b}), we can estimate that the birth-rate of
pulsars with magnetic fields $B \sim 6\times10^{11}$~G is a fraction
0.05 of that of pulsars with magnetic fields comparable to the Crab Pulsar.
Thus of the $\sim$35 Crab-like and composite SNRs known in the Galaxy,
we can expect $\sim$2 to be powered by such low-field pulsars, and \g328\
and G74.9+1.2 may thus represent the complete sample of such sources.

The radio beaming fraction for young pulsars is estimated to be in
the range 50--70\% (\cite{fm93}; \cite{bj98}).  Thus the failure to
detect radio pulsations from \g328, G74.9+1.2 and N157B (\cite{kmj+96};
\cite{llc98}; \cite{ckm+98}) is unlikely to be solely due to beaming,
and more likely results from the large distance to these objects. With
continuing improvements in sensitivity, searches for pulsations towards
these sources should eventually be successful.

\begin{acknowledgements}

We thank Taisheng Ye for early assistance with this project, and Pat
Slane and Jack Hughes for useful discussions on the manuscript. We also
acknowledge some important suggestions made by the referee, Roger
Chevalier.  
The Australia Telescope is funded by the Commonwealth of Australia for
operation as a National Facility managed by CSIRO.  B.M.G.
acknowledges the support of NASA through Hubble Fellowship grant
HF-01107.01-98A awarded by the Space Telescope Science Institute, which
is operated by the Association of Universities for Research in
Astronomy, Inc., for NASA under contract NAS 5--26555. J.R.D.
acknowledges a Visitor's Fellowship from the Netherlands Scientific
Organization (NWO) during his stay at the NFRA.

\end{acknowledgements}


\bibliographystyle{apj1}
\bibliography{journals,modrefs,psrrefs,crossrefs}


\vspace{2cm}


\clearpage

\begin{table}[hbt]
\caption{ATCA observations of \g328.}
\begin{tabular}{lccc} \hline \hline
                    & 1.4~GHz   & 4.5~GHz \\ \hline
Dates Observed      & 1999 July 31 & 
\multicolumn{1}{l}{1993 May 23, 1993 July 19,} \\
                &    & \multicolumn{1}{l}{1993 Sep 04, 1995 Jan 22,} \\
		&    & \multicolumn{1}{l}{1995 Feb 26} \\
Resolution  & $19\farcs5 \times 16\farcs5$ & $2\farcs0 \times 1\farcs5$ \\
rms noise (mJy~beam$^{-1}$) & 0.8 & 0.04 & (Stokes $I$) \\
                                & 0.1 & 0.04 & (Stokes $V$) \\ 
Flux density of \g328\ (Jy)  & $14.3\pm0.1$ & $12.5\pm0.2$ \\ \hline
Center of \g328\ & \multicolumn{3}{l}{$15^{\rm h}55^{\rm m}33^{\rm s}$, 
$-53\arcdeg17\arcmin00\arcsec$
 ($\alpha$, $\delta$; J2000)} \\
   & \multicolumn{3}{l}{328.42 +0.22 ($l$, $b$)} \\
Diameter of \g328\  & \multicolumn{2}{l}{$5\farcm0 \times 5\farcm0$} \\ \hline
\end{tabular}
\label{tab_obs}
\end{table}

\vspace{2cm}

\begin{table}[hbt]
\caption{Properties of \g328\ and two other Crab-like SNRs of comparable
radio luminosities and diameters. The Crab Nebula is included for
comparison.} 
\begin{tabular}{lccccccc} \hline \hline
SNR        &    $L_R$\tablenotemark{1}  & Diameter  & 
\multicolumn{4}{c}{Properties of associated pulsar\tablenotemark{2}} 
& Ref\tablenotemark{3} \\
     &   ($10^{35}$~erg~s$^{-1}$) & (pc) &  $P$~(ms) & $\tau_c$ (kyr)
    &  $\dot{E}$~($10^{38}$~erg~s$^{-1})$ & $B$ ($10^{12}$~G) \\ \hline
\g328      & 3.3 & 25 & 11 & 7 & 8.3 & 0.6 &  1 \\
G74.9+1.2      & 0.7 & 25 &  18 & 11 & 1.8 & 0.7  &  2  \\
N157B (observed)   &  3.5   & 24   & 16 & 5 & 4.8 & 0.9 &  3, 4 \\
N157B (inferred)   & ''   & '' &  11 & 6 & 8.8 & 0.6 \\ \hline
Crab Nebula & 1.8 & 3.5 &  33 & 0.9 & 4.5 & 3.8 &  2, 5 \\ \hline
\end{tabular}

\tablenotetext{1}{Integrated between 100~MHz and 100~GHz.}
\tablenotetext{2}{No pulsar has yet been detected in \g328\ or G74.9+1.2, and
the properties given here have been inferred using the method described
in the text. For N157B, both inferred and actual pulsar properties are
listed.}
\tablenotetext{3}{References: (1) this paper; (2) Helfand \& Becker
(1987)\nocite{hb87}; (3) Marshall \etal\ (1998)\nocite{mgz+98};
(4) Lazendic \etal\ (2000\nocite{ldh+00}); (5) Manchester \& Taylor
(1977\nocite{mt77})}
\label{tab_compare}
\end{table}

\end{document}